\documentclass[aps,reprint,showpacs,showkeys,prl,groupedaddress]{revtex4-1}

\usepackage{caption}
\usepackage{amsmath}
\usepackage{graphics}
\usepackage{graphicx}
\begin{document}



\title{Polarised Drell-Yan results from COMPASS}


\author{
Celso Franco\\ 
on behalf of the COMPASS collaboration\\
~\\
}
\affiliation{LIP - Laborat\'orio de Instrumenta\c{c}\~ao e F\'\i sica Exper\'\i mental de Part\'\i culas\\ Av. Prof. Gama Pinto, n.2, Complexo Interdisciplinar (3is),
1649-003 Lisboa, Portugal }
 


\begin{abstract}
The COMPASS experiment at CERN is one of the leading experiments studying the nucleon spin
structure. Until 2012 the Parton Distribution Functions and the Transverse Momentum Dependent
Parton Distribution Functions (TMDs) were extensively studied at COMPASS using Semi-Inclusive
Deep Inelastic Scattering measurements. In 2015, the Drell-Yan measurements with a negative pion
beam interacting with a transversely polarized ammonia target have started and will be continued
through 2018. The goal is to access the TMDs of both pions and protons without any prior knowledge
about fragmentation functions. Since the Drell-Yan data cover the same kinematic region of the 
semi-inclusive data, COMPASS has the unique opportunity to test the sign change of the Sivers TMD as
predicted by QCD. In this article the first measurement of spin dependent azimuthal asymmetries in
the pion induced Drell-Yan process will be presented. These asymmetries, which are related to the
convolution of pion and nucleon TMDs, are extracted from pairs of oppositely charged muons with
invariant masses between 4.3 GeV/c$^{2}$ and 8.5 GeV/c$^{2}$
\end{abstract}


\maketitle
\onecolumngrid

\section{Introduction}

The COMPASS experiment at CERN studies TMDs using polarised Drell-Yan and polarised semi-inclusive DIS measurements. 
The Drell-Yan process consists in an electromagnetic annihilation of a quark-antiquark pair with the production of 
two leptons in the final state. COMPASS uses the dimuon channel, to take advantage of the excellent muon detection 
by the spectrometer\cite{COMPASS}, with the goal of studying the TMDs of the valence up quark in the proton. This goal is  
accomplished by impinging a negative pion beam of 190 GeV/c in a transversely polarised proton target (cf. 
left-panel of Fig.~\ref{DY_process}):

\vspace{0.3 cm}

\begin{figure}[h]
\includegraphics[width=0.65\columnwidth]{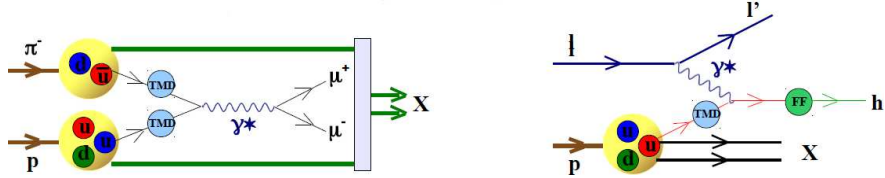}
\caption{The Drell-Yan (left) and SIDIS (right) processes\label{DY_process}}
\end{figure}

\vspace{0.3 cm}

 The need of TMDs to properly describe the nucleon structure became evident from 
the results of two unpolarised Drell-Yan experiments from the 1980s: NA10\cite{NA10} and E615\cite{E615}. By studying 
the angular dependence of the Drell-Yan cross section

\begin{equation}
\frac{1}{\sigma}\frac{d\sigma}{d\Omega} = \frac{3}{4\pi}\frac{1}{\lambda+3}[1+\lambda \cos^{2}\theta + \eta \sin 2\theta \cos \phi +\frac{\nu}{2}\sin^{2}\theta \cos2\phi]
\end{equation}

\begin{figure}[h]
\includegraphics[width=0.32\columnwidth]{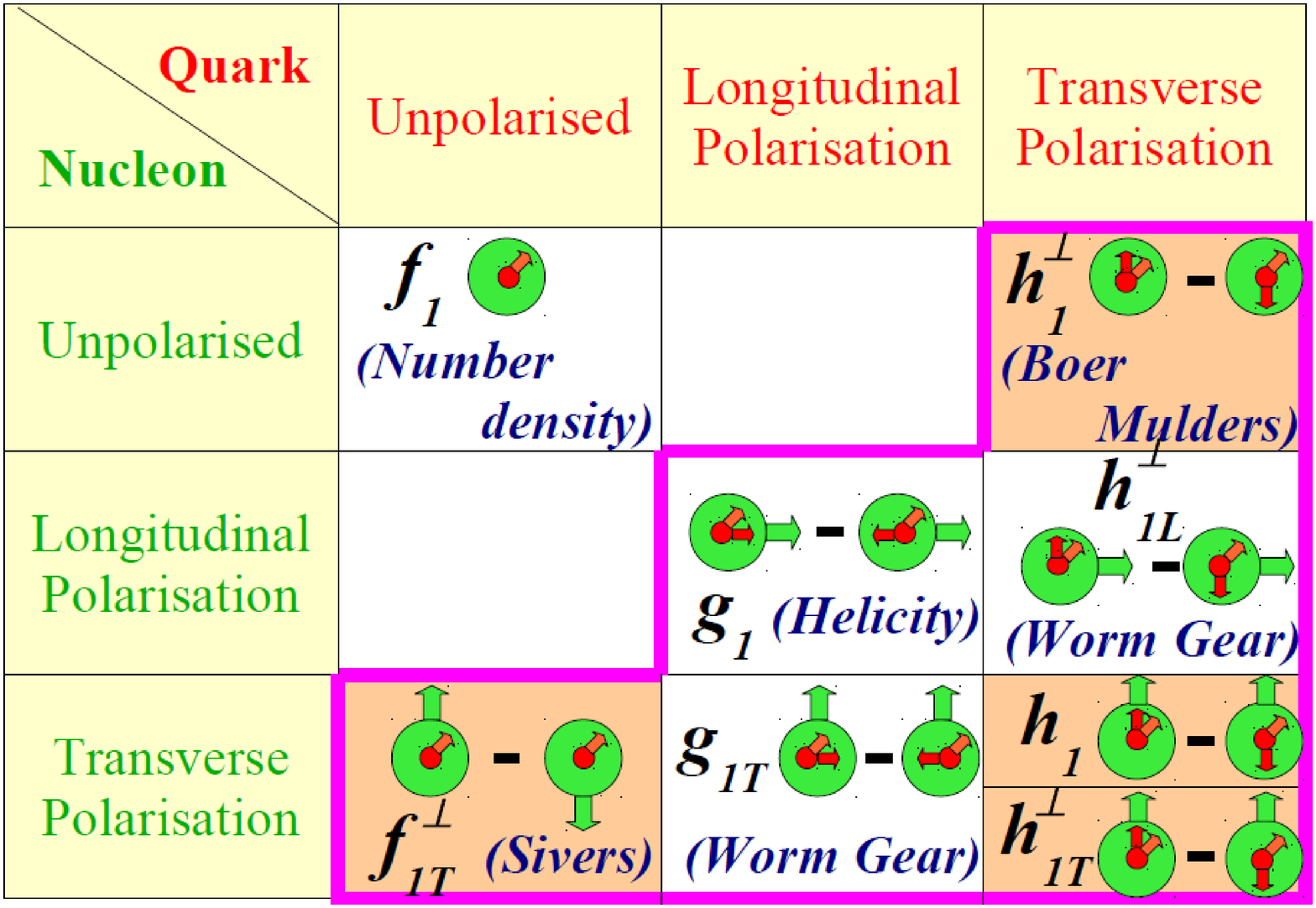}
\caption{Proton structure in the momentum space at Twist-2. The $h_{1}(x,k_{T})$ and $h_{1T}^{\bot}(x,k_{T})$ TMDs are the so-called Transversity and Pretzelosity functions. The 3 collinear PDFs $f_{1}(x)$, $g_{1}(x)$ and $h_{1}(x)$ are the only ones surviving an integration over $k_{T}$ (orange arrow). \label{TMDs}}
\end{figure}

\vspace{0.3 cm}

\noindent where $\theta$ and $\phi$ are the polar and azimuthal angles of one of the leptons in the dilepton rest 
frame, those experiments measured a $\cos 2\phi$ modulation up to 30\%. This striking result gave rise to several 
interpretations. One conclusion was that the intrinsic transverse momentum $k_{T}$ of quarks is not a negligible 
quantity inside the protons ($\phi$ modulations are not expected for $k_{T} = 0$).  Consequently, instead of 
three collinear PDFs such as the unpolarised $f_{1}$, the helicity $g_{1}$ and the transversity $h_{1}$, a total of 8 TMDs 
are needed to describe the nucleon structure at leading-twist (cf. Fig.~\ref{TMDs}). 

The key feature of the Drell-Yan process, when compared with the SIDIS process (cf. right-panel of Fig.~\ref{DY_process}), is the 
possibility to study a proton TMD convoluted with a pion TMD. In the SIDIS case one needs to know the Fragmentation 
Function (FF) of the struck quark instead of a pion TMD. Regarding the Drell-Yan experiment, COMPASS is able to study the 
Sivers, Transversity and Pretzelosity TMDs, together with two higher-twist TMDs, by scattering 190 GeV/c pions off a transversely 
polarised proton target. The T-odd Sivers function, which describes the transverse motion of quarks induced by the transverse spin 
of the nucleon, is of extreme importance because it contains information about the orbital angular momentum (OAM) of quarks. 
However, the main motivation for a detailed study of the Sivers TMD is the QCD prediction that this TMD must change sign 
when accessed via Drell-Yan or SIDIS processes \cite{Collins}. This prediction is a crucial test to our current understanding of 
TMDs: the mentioned sign change ensures the TMD factorisation. Currently COMPASS is the only experiment capable of performing both 
SIDIS and Drell-Yan measurements with the same spectrometer (using the virtual-photon channel).

\section{The Drell-Yan and SIDIS Asymmetries}

The Drell-Yan and SIDIS cross-sections are expressed in terms of azimuthal spin asymmetries. The amplitude of each angular 
modulation contains a pion TMD convoluted with proton TMD, in the Drell-Yan case, and a proton TMD convoluted with 
a quark FF in the SIDIS case. At leading-twist, the Drell-Yan cross-section can be written as 

\vspace{-0.1cm}

\begin{equation}
\begin{split}
d\sigma^{DY}_{Twist-2} \propto ~&(1 + cos^{2}(\theta) + sin^{2}(\theta)A_{UU}^{cos(2\phi)}cos(2\phi) + S_{T}[(1 + cos^{2}(\theta))A_{UT}^{sin(\phi_{S})}sin(\phi_{S})\\ &+ sin^{2}(\theta)(A_{UT}^{sin(2\phi - \phi_{S})}sin(2\phi - \phi_{S}) + A_{UT}^{sin(2\phi + \phi_{S})}sin(2\phi + \phi_{S}))])
\end{split}
\end{equation}

\noindent and the SIDIS cross-section as

\begin{figure}[h]
\includegraphics[width=0.62\columnwidth]{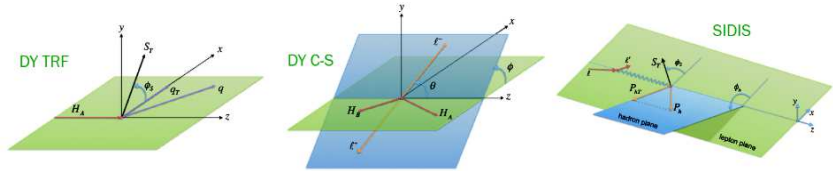}
\caption{Azimuthal and polar angles for Drell-Yan (left and middle) and SIDIS (right).\label{Frames}}
\end{figure}

\vspace{-0.1cm}

\begin{equation}
\begin{split}
d\sigma^{SIDIS}_{Twist-2} \propto ~&(1 + \epsilon cos(2\phi_{h})A_{UU}^{cos(2\phi_{h})}  + S_{T}[sin(\phi_{h} - \phi_{S})A_{UT}^{sin(\phi_{h} - \phi_{S})}\\ &+ \epsilon sin(\phi_{h} + \phi_{S})A_{UT}^{sin(\phi_{h} + \phi_{S})} +  \epsilon sin(3\phi_{h} + \phi_{S})A_{UT}^{sin(3\phi_{h} + \phi_{S})}])
\end{split}
\end{equation}

\noindent where the subscripts in the asymmetries represent the  beam and target polarisations: $U$ stands for 
unpolarised and $T$ for transversely polarised. $S_{T}$ represents the direction of the target spin, $\epsilon = (1 - y -1/4\gamma^{2}y^{2})/(1 - y + 1/2y^{2} + 1/4\gamma^{2}y^{2})$ and $\gamma = 2Mx/Q$. The interpretation for the polar and azimuthal angles is provided by Fig.~\ref{Frames}

\noindent and the TMD interpretation of the azimuthal spin asymmetries can be seen in Table~\ref{table}.

\begin{table}[]
\begin{tabular}{| c | c | c |}
\hline
\textbf{Drell-Yan} & \textbf{Proton TMDs} & \textbf{SIDIS}\\
\hline\hline
$A_{UU}^{cos(2\phi)} \propto h_{1,\pi}^{\bot q} \bigotimes h_{1,p}^{\bot q}$ & Boer-Mulders & $A_{UU}^{cos(2\phi_{h})} \propto h_{1,p}^{\bot q} \bigotimes H_{1q}^{\bot h}$\\
\hline
$A_{UT}^{sin(\phi_{S})} \propto f_{1,\pi}^{q} \bigotimes f_{1T,p}^{\bot q}$ & Sivers & $A_{UT}^{sin(\phi_{h} - \phi_{S})} \propto f_{1T,p}^{\bot q} \bigotimes D_{1q}^{h}$\\
\hline
$A_{UT}^{sin(2\phi - \phi_{S})} \propto h_{1,\pi}^{\bot q} \bigotimes h_{1,p}^{q}$ & Transversity & $A_{UT}^{sin(\phi_{h} + \phi_{S})} \propto h_{1,p}^{q} \bigotimes H_{1q}^{\bot h}$\\
\hline
$A_{UT}^{sin(2\phi + \phi_{S})} \propto h_{1,\pi}^{\bot q} \bigotimes h_{1T,p}^{\bot q}$ & Pretzelosity & $A_{UT}^{sin(3\phi_{h} - \phi_{S})} \propto h_{1T,p}^{\bot q} \bigotimes H_{1q}^{\bot h}$\\
\hline
\end{tabular}
\caption{Interpretation of the leading-twist Asymmetries. The proton TMDs are convoluted with pion TMDs, when accessed via Drell-Yan, and convoluted with a spin-(in)dependent quark FF ($D$)$H$ when accessed via SIDIS\label{table}}
\end{table}

\section{Extraction of the Transverse Spin Asymmetries}

A total of 35000 Drell-Yan dimuons, with invariant masses between 4.3 GeV/c$^{2}$ and 8.5 GeV/c$^{2}$, were used for the extraction of the 
TSAs (Transverse Spin Asymmetries). The final dimuon mass spectrum, after all the quality cuts, is shown in Fig.~\ref{Mass}:

\vspace{0.4 cm}

\begin{figure}[h]
\includegraphics[width=0.5\columnwidth]{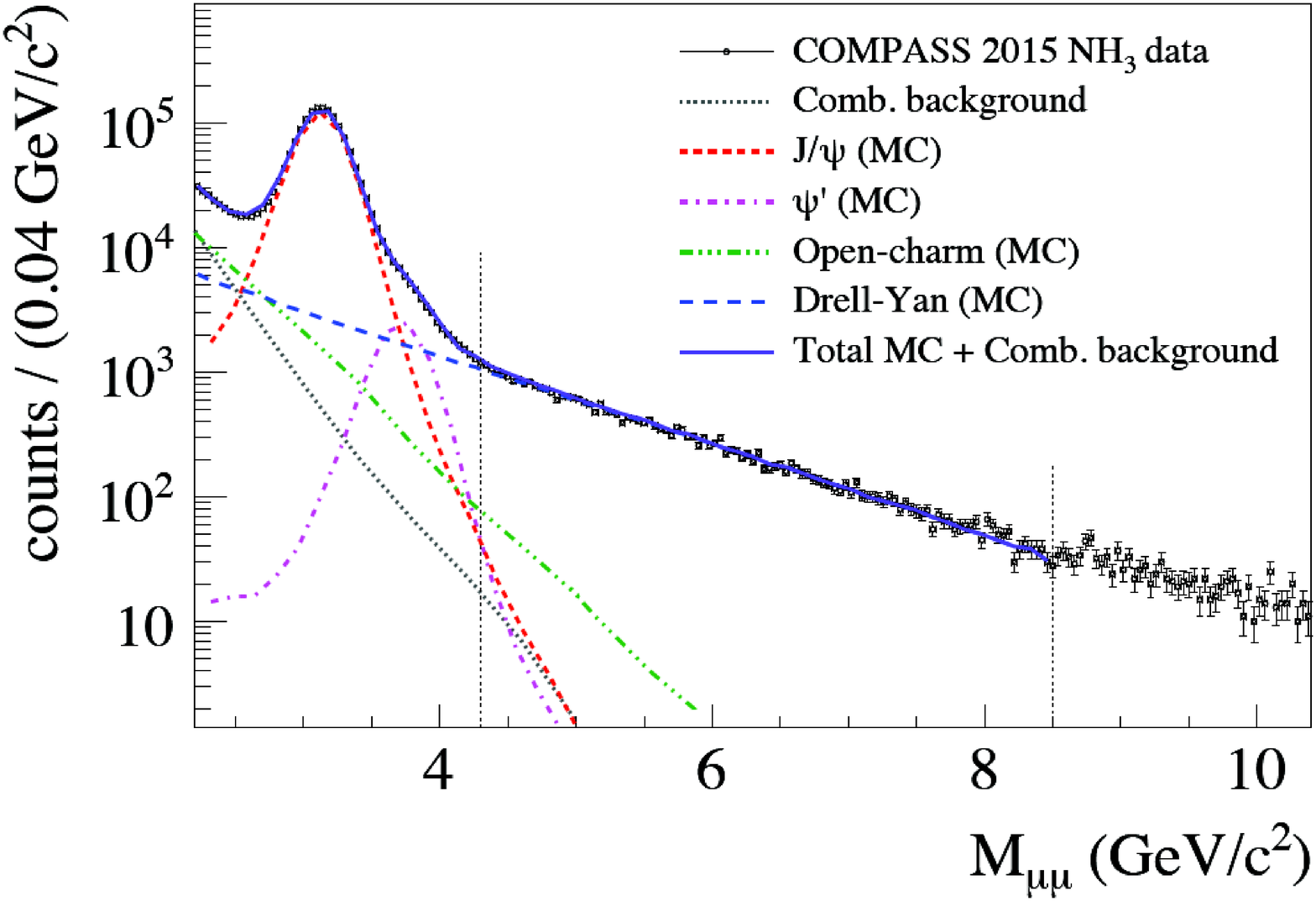}
\caption{The Drell-Yan dimuon mass spectrum. Good agreement between Data and Monte-Carlo is observed\label{Mass}}
\end{figure}

\vspace{0.4 cm}

\indent The mass window for the analysis was defined to maximize the Drell-Yan statistics while keeping the contamination by other 
processes under control. In the specified mass window the contamination is smaller than 4\%.\\ 
\\
\indent A total of 3 Twist-2 and 2 Twist-3 TSAs were simultaneously extracted from the Drell-Yan data using an unbinned maximum 
likelihood method \cite{likelihood}. The general expression for the Drell-Yan cross section, accounting for all orders, can be 
written as follows

\begin{equation}
\begin{split}
d\sigma^{DY} \propto ~&(1 + D_{[sin(2\theta)]}A_{UU}^{cos(\phi)}cos(\phi) +  D_{[sin^{2}(\theta)]}A_{UU}^{cos(2\phi)}cos(2\phi) +  S_{T}[D_{[1 + cos^{2}(\theta)]}A_{UT}^{sin(\phi_{S})}sin_(\phi_{S})\\ 
&+ D_{[sin^{2}(\theta)]}(A_{UT}^{sin(2\phi - \phi_{S})}sin_(2\phi - \phi_{S}) + A_{UT}^{sin(2\phi + \phi_{S})}sin_(2\phi + \phi_{S}))\\ 
&+ D_{[sin(2\theta)]}(A_{UT}^{sin(\phi - \phi_{S})}sin_(\phi - \phi_{S}) + A_{UT}^{sin(\phi + \phi_{S})}sin_(\phi + \phi_{S}))])
\end{split}
\end{equation}

\vspace{0.4 cm}

\noindent where $D$ represent the different Depolarisation factors weighting the azimuthal asymmetries. The experimental asymmetries, 
which are related to the physical azimuthal asymmetries as $A^{raw} = P_{T}fDA_{UT}$, are determined by counting the number of dimuons 
coming from each spin configuration of the target. The factors $P_{T}$ and $f$ represent the target polarisation and the fraction of 
polarisable material inside the target (dilution factor). The target is composed by two 55 cm long cylindrical cells, separated by a 
20 cm gap, having each one a radius of 2 cm and being filled with $NH_{3}$ material. The protons from both cells are transversely 
polarised in opposite directions to allow for the extraction of the spin asymmetries. The differences of acceptance between cells 
are canceled out by periodically reversing the polarisation direction for both cells. Using a total of 18 spin configurations, 
resulting from 9 periods of data (each period contains data taken with very similar experimental conditions) subdivided in 2 sub-periods
 where the cells polarisations are reversed, one is able to fit in an optimal way all the 7 asymmetries of Eq. 4. For the extraction 
of the TSAs the depolarisation factors

\begin{equation}
\begin{split}
D_{[sin(2\theta)]} = \frac{sin(2\theta)}{1 + \lambda cos^{2}(\theta)},~~~ D_{[1 + cos^{2}(\theta)]} = \frac{1 + cos^{2}(\theta)}{1 + \lambda cos^{2}(\theta)},~~~ D_{[sin^{2}(\theta)]} = \frac{sin^{2}(\theta)}{1 + \lambda cos^{2}(\theta)}
\end{split}
\end{equation}

\noindent where determined for $\lambda = 1$. Such value for the polar angle asymmetry imply a $k_{T} = 0$ 
(naive Drell-Yan scenario). In order to account for other scenarios with $\lambda \neq 1$, a maximum uncertainty of $5\%$ was assigned to 
the determination of the depolarisation factors (cf. Fig.~\ref{Dep}):

\vspace{0.4cm}

\begin{figure}[h]
\includegraphics[width=0.47\columnwidth]{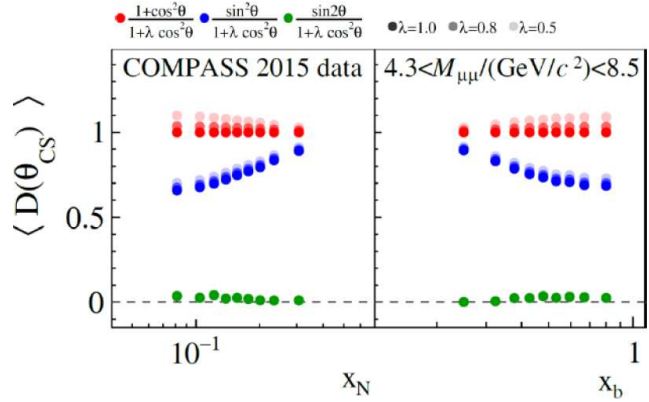}
\caption{Depolarisation factors for three values of $\lambda$ as a function of $x_{N}$ and $x_{b}$\label{Dep}}
\end{figure}

\vspace{0.4cm}

\indent The asymmetries were also weighted with the dilution factor $f$ to optimize the statistical presicion of the results. 
This factor was determined from Monte-Carlo, event-by-event, having an average value of about $18\%$ and 
an uncertainty of $8\%$. This factor accounts also for the small migration of events from one target cell to the other. The final 
asymmetries were also corrected by the average value of the target polarisation which amounts to $P_{T} \approx 73\%$. The obtained 5 TSAs 
are shown in Fig.~\ref{Asymmetries}:

\vspace{0.4cm}

\begin{figure}[h]
\includegraphics[width=0.9\columnwidth]{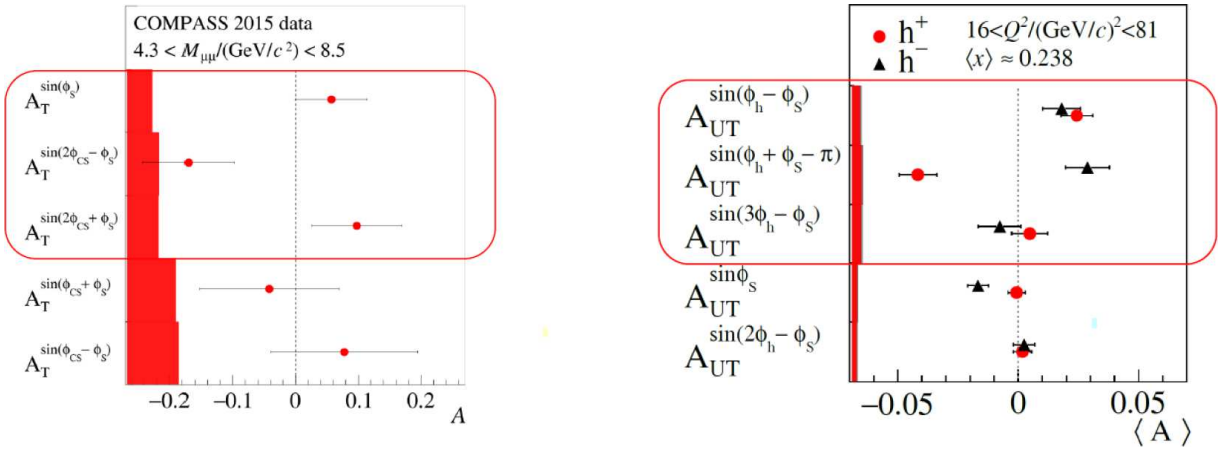}
\caption{TSAs obtained from Drell-Yan (left) and SIDIS (right) processes. The first 3 are Twist-2 asymmetries encoding information about 
the Sivers, Transversity and Pretzelosity TMDs.\label{Asymmetries}}
\end{figure}

\vspace{0.4cm}

\indent The same TSAs obtained from a SIDIS measurement are also shown in Fig.~\ref{Asymmetries} for comparison. Both results are 
published \cite{DY, SIDIS}. In Fig.~\ref{Kinematics} one can confirm that the high $Q^{2}$ SIDIS data cover the same kinematic region 
of the Drell-Yan data:  

\vspace{0.4cm}

\begin{figure}[h]
\includegraphics[width=0.85\columnwidth]{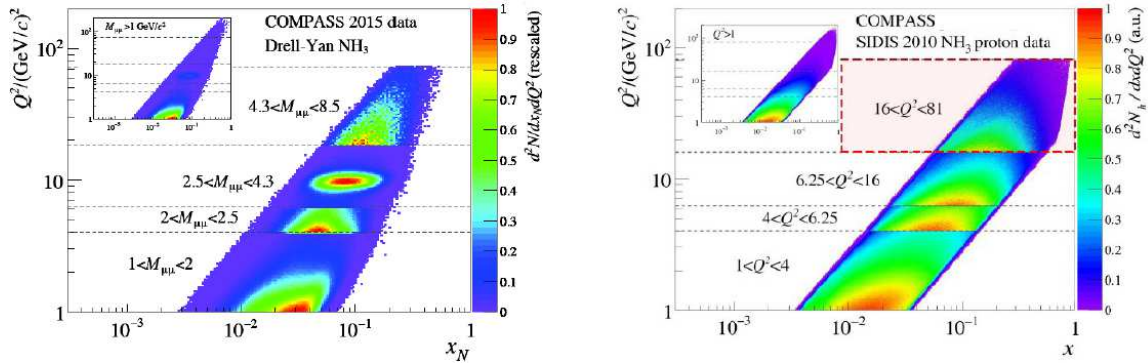}
\caption{$Q^{2}$ vs $x$ for Drell-Yan (left) and SIDIS (right). \label{Kinematics}}
\end{figure}

\vspace{0.2cm}

\begin{figure}[h]
\includegraphics[width=0.55\columnwidth]{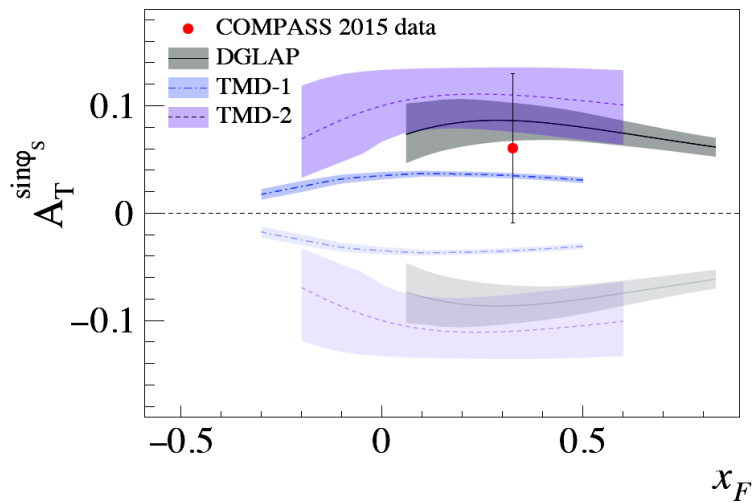}
\caption{The Drell-Yan Sivers asymmetry. Taking as a reference the COMPASS SIDIS measurement for the Sivers asymmetry, the sign change 
scenario is verified only if the Drell-Yan Sivers asymmetry is found to be positive. The DGLAP model is a fit to the COMPASS, HERMES 
and JLab SIDIS data \cite{DGLAP}. The TMD-1 model is a fit to the COMPASS, HERMES and JLab SIDIS data using a $p_{T} < 1$ GeV/c cut and 
invariant masses between 4 and 9 GeV/c$^{2}$ \cite{TMD1}. The TMD-2 model differs only by the $p_{T} < 2$ GeV/c cut \cite{TMD2}. \label{Sivers}}
\end{figure}

\noindent Therefore, a direct comparison of the Sivers asymmetry from both processes is straightforward. Presently, the 
sign change scenario is favored by the COMPASS data with a significance of $1\sigma$ (cf. Fig.~\ref{Sivers}) 

\vspace{0.2cm}

\subsection{}
\subsubsection{}

\section{Conclusions}

In 2015 COMPASS performed the first ever polarised Drell-Yan measurement using a pion beam. The 
Pretzelosity asymmetry was found to be positive with a significance of $1\sigma$. The 2 Twist-3 
asymmetries are also compatible with zero within $0.5\sigma$. The Transversity asymmetry is 
negative with a significance of $2\sigma$, and the Sivers asymmetry was measured with the right 
sign to fulfill the QCD prediction of sign change when compared with the COMPASS SIDIS measurement.
The latter is  more than $3\sigma$ away from zero, implying a relevant contribution of the OAM of valence quarks 
to the nucleon spin.\\
\indent The COMPASS Drell-Yan measurement provided the first world test to the universality of the T-Odd TMDs 
using the virtual-photon channel for the lepton pair production. A second year of polarised Drell-Yan 
data taking will take place in 2018 to improve the experimental uncertainties. Currently the Drell-Yan Sivers 
asymmetry has only a significance of $1\sigma$. The new data, together with a possible enlargement 
of the dimuon mass window (using machine learning techniques to separate processes in a multidimensional 
way), will surely contribute to improve the confidence level of the sign change scenario. 

\section{Acknowledgments}
This work was supported by the Portuguese Funda\c{c}\~ao para a Ci\^encia e a Tecnologia (CERN/FIS-PAR/0007/2017)


\begin{thebibliography}{99}


\bibitem{COMPASS}
COMPASS collaboration, NIMA 577 (2007) 455-518, arXiv: 0703049 [hep-ex]
\bibitem{NA10}
NA10 Collaboration, S. Falciano et al, Z. Phys. C 31 (1986) 513
\bibitem{E615}
E615 Collaboration, J. S. Conway et al, Phys. Rev. D 39 (1989) 92
\bibitem{Collins}
J.C. Collins, Phys. Lett. B536 (2002) 43
\bibitem{likelihood}
COMPASS collaboration, Phys. Lett.B713, 10 (2012), arXiv:1202.6150 [hep-ex]
\bibitem{DY}
COMPASS collaboration, PRL 119 (2017) 112002, arXiv:1704.00488 [hep-ex]
\bibitem{SIDIS}
COMPASS collaboration, PBL 770 (2017) 138, arXiv:1609.07374 [hep-ex]
\bibitem{DGLAP}
M. Anselmino et al., arXiv:1612.06413 [hep-ex]
\bibitem{TMD1}
M. Echevarria et al., PRD 89 (2014) 074013
\bibitem{TMD2}
P. Sun and F. Yuan, PRD 88 (2013) 114012

\end{thebibliography}
\end{document}